\documentclass{aa}  

\usepackage{graphicx}
\usepackage{txfonts}
\usepackage[switch]{lineno}
\bibpunct{(}{)}{;}{a}{}{,}
\usepackage{float}
\usepackage{hyperref}
\usepackage{natbib}
\hypersetup{
    colorlinks=true,
    linkcolor=blue,
    filecolor=magenta,      
    urlcolor=blue,
    citecolor=blue,
}
\begin{document}

   \title{$\eta$ Carinae with \textit{Fermi}-LAT: \\Two full orbits and the third periastron}
   \titlerunning{$\eta$ Carinae and nearby systems with \textit{Fermi}-LAT}

   \author{G. Martí-Devesa
          \and
          O. Reimer
          }
	\authorrunning{Martí-Devesa \& Reimer}
   \institute{Institut f\"{u}r Astro- und Teilchenphysik, Leopold-Franzens-Universit\"{a}t Innsbruck, A-6020 Innsbruck, Austria\\\email{guillem.marti-devesa@uibk.ac.at}
}

   \date{Received ---, ---; accepted ---, ---}

  \abstract
   {Colliding-wind binaries are massive stellar systems featuring strong, interacting winds. These binaries may be actual particle accelerators, making them variable $\gamma$-ray sources due to changes in the wind collision region along the orbit. However, only two of these massive stellar binary systems have been identified as high-energy sources. The first and archetypical system of this class is $\eta$ Carinae, a bright $\gamma$-ray source with orbital variability peaking around its periastron passage.}
   {The origin of the high energy emission in $\eta$ Carinae is still unclear, with both lepto-hadronic and hadronic scenarios being under discussion. Moreover, the $\gamma$-ray emission seemed to differ between the two periastrons previously observed with the \textit{Fermi} Large Area Telescope. Continuing observations might provide highly valuable information for the understanding of the emission mechanisms in this system.}
   {We have used almost 12 years of data from the \textit{Fermi} Large Area Telescope. We studied both low and high energy components, searching for differences and similarities between both orbits, and made use of this large dataset to search for emission from nearby colliding-wind binaries.}
   {We show how the energy component above 10 GeV of $\eta$ Carinae peaks months before the 2014 periastron, while the 2020 periastron is the brightest to date. Additionally, upper limits are provided for the high-energy emission in other particle-accelerating colliding-wind systems.}
   {Current $\gamma$-ray observations of $\eta$ Carinae strongly suggest that the wind collision region of this system is perturbed from orbit to orbit affecting particle transport within the shock.}


   \maketitle
%

\section{Introduction} \label{sec:intro}

Binary systems are natural environments to study shocks under variable, periodical conditions. An emerging class of these systems is colliding-wind binaries (CWBs), massive stellar systems with powerful stellar winds \citep{Becker13}. Such strong winds eventually interact, forming a bow-shocked wind collision region (WCR) delimited by two separated shock fronts surrounding the star with a weaker wind \citep{Eichler93}. Diffusive shock acceleration (DSA) can occur under these circumstances, accelerating particles up to high energies and leading to the emission of non-thermal radiation \citep{Benaglia03, Reimer06, Becker07, Reitberger14em, Grimaldo19, Pittard20}.

One of the most luminous and intriguing known Galactic sources is \object{$\eta$ Carinae}, which has been monitored at different wavelengths for decades. It is a CWB with a primary Luminous Blue Variable (LBV) star with $M_{\eta Car_A} \geq 90$ M$_{\sun}$ \citep{Hillier01}. Its companion has not been directly observed, but inferred from its orbital variability to be an O or Wolf Rayet (WR) star with $M_{\eta Car_B}\sim 30-50$ M$_{\sun}$ \citep{Hillier01,Mehner10-Components}. $\eta$ Car is located in the Carina arm at a distance of $2350 \pm 50$ pc \citep{Smith06}, surrounded by the Homunculus nebula. For a detailed review on the history and characteristics of this system and its surroundings, see \cite{Davidson12}. The central binary system has a period of $P \sim 2024$ days \citep[periastron at $T_0=50799.3$ MJD,][]{Corcoran05}, in a highly eccentric orbit of $e\sim0.9$ \citep{Nielsen07}. The powerful winds of both components have large mass-loss rates of $ \dot{M}_{\eta Car_A} \sim 2.5 \cdot 10^{-4}$ $M_{\sun}$ yr$^{-1}$ and $ \dot{M}_{\eta Car_B} \sim 10^{-5}$ $M_{\sun}$ yr$^{-1}$, with terminal velocities of $500$ km s$^{-1}$ and $3000$ km s$^{-1}$, respectively \citep{Pittard02}.

Although no radio synchrotron emission has been detected from $\eta$ Carinae \citep{Duncan03}, it is a non-thermal X-ray \citep{Leyder08, Sekiguchi09, Hamaguchi18} and $\gamma$-ray \citep{Tavani09, Abdo10, Reitberger15, HESS20} emitter. Its spectrum has been widely studied above 100 MeV, where two distinct components are detected above (High Energy; HE) and below (Low Energy; LE) 10 GeV. Despite the consensus that the HE component has a hadronic origin, at LE the situation is unclear: both leptonic \citep{Farnier11, Gupta17} and hadronic \citep{Ohm15, White20} scenarios are still plausible.

Unfortunately, only one other CWB has been detected in $\gamma$-rays: $\gamma ^2$ Velorum, also known as WR~11 \citep{Pshirkov16, Marti-Devesa20}. Despite numerous efforts, CWBs with significant synchrotron radio emission have been elusive to detection both with orbital and ground-based $\gamma$-ray observatories \citep{Romero99, Aliu08, Werner13, Palacio20}. In particular, the upper limits reported in observations on \object{WR~140}, \object{WR~146} or \object{WR~147} \citep{Werner13} imply very low efficiencies for $\gamma$-ray emission via the Inverse Compton (IC) process in classical CWBs. Therefore, the best target source to understand CWBs is the bright $\eta$ Carinae itself.

\begin{figure*}[h]
   \centering
   \includegraphics[width=2.0\columnwidth]{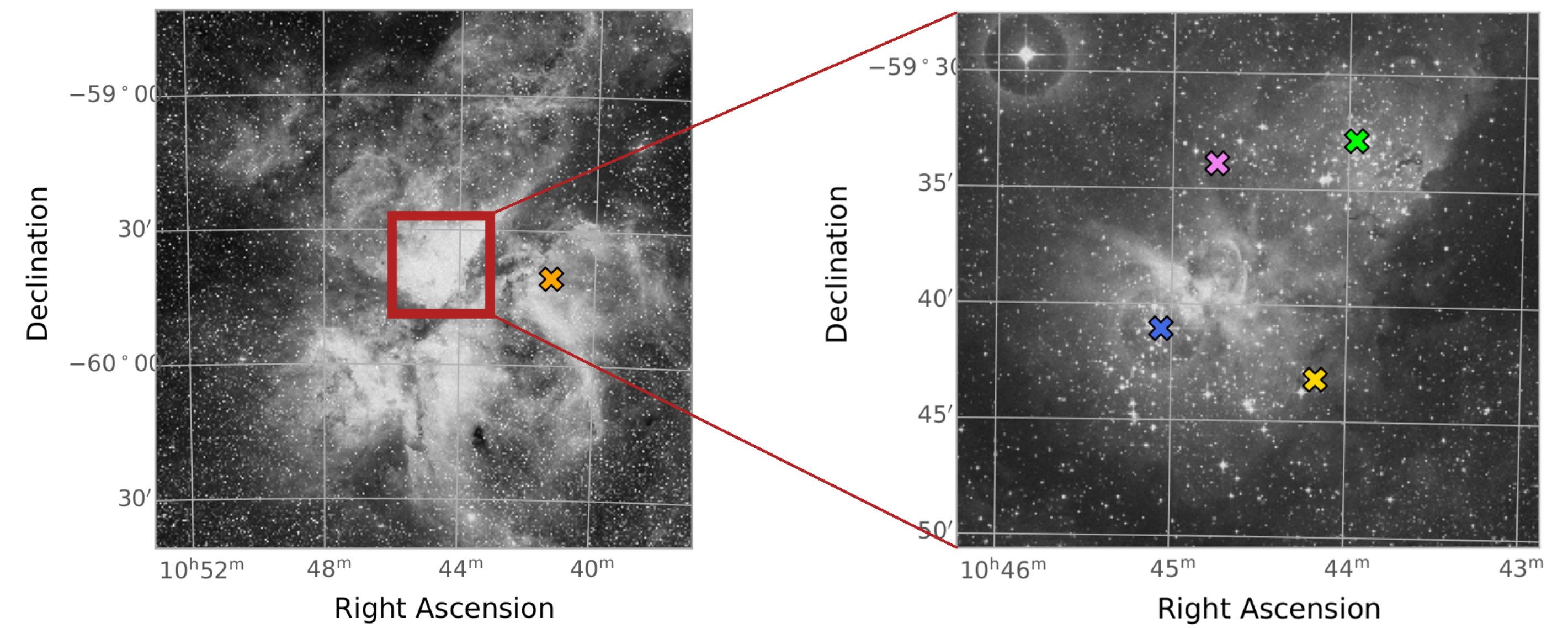}
   \caption{\textit{Left}: Carina Nebula as seen in the Second Generation Digitalized Sky Survey (DSS2; red filter), which contains the system WR 22 (orange). In red, zoomed region. \textit{Right}: Core of the Carina Nebula centred on $\eta$ Carinae as seen in DSS2 (IR filter). The region shown contains $\eta$ Carinae (blue), WR 25 (yellow), HD 93129A (green) and HD 93250 (pink), while WR 21a and WR 39 lie $\sim 3^{\circ}$ away from $\eta$ Carinae, outside the nebula.}
     
         \label{FigVibStab}
   \end{figure*}

In this work, we perform a detailed study on the two full orbits of $\eta$ Carinae observed by the $Fermi$-LAT in 12 years of operations, including three periastron passages. Additionally, we take advantage of this data selection to search for new nearby $\gamma$-ray source candidates, with special interest in other CWBs (Figure \ref{FigVibStab}). Section \ref{sec:analysis} presents the analysis performed, and results are shown in Section \ref{sec:results}. Then, those are discussed in Section \ref{sec:discussion}. Finally, the conclusions of our study are summarised in Section \ref{sec:summary}. 

\section{Observations and Analysis} \label{sec:analysis}


The Large Area Telescope (LAT) is the main instrument on board the \textit{Fermi Gamma-ray Space Telescope} \citep{FermiLAT}, covering the energy range between 30 MeV to more than 100 GeV. Its energy-dependent point-spread function (PSF) goes from more than $5^{\circ}$ below 100 MeV to less than $0.1^{\circ}$ above 10 GeV at $68\%$ containment. In this paper, observations from 2008 August 4 to 2020 May 29 are included. The analysis was performed using {\it Fermitools-1.2.23}\footnote{This is the nomenclature for the \textit{Fermi} Science Tools released through Conda. See \url{https://github.com/fermi-lat/Fermitools-conda/wiki}} on P8R3 data \citep{Atwood13, P8R3}. Fluxes are obtained performing a binned maximum likelihood fit \citep{Mattox96} using \texttt{fermipy 0.19}\footnote{Python package for the $Fermitools$. See \url{https://fermipy.readthedocs.io/en/latest/}} \citep{Fermipy}. We used the 4FGL DR2 catalogue (\texttt{gll\_psc\_v23}) for our source model \citep{4FGL, Ballet20}, while the diffuse emission was assessed using 'gll$\_$iem$\_$v07.fits' and 'iso$\_$P8R3$\_$SOURCE$\_$V2$\_$v1.txt' for the Galactic and isotropic components, respectively\footnote{The latest background models are provided in \url{https://fermi.gsfc.nasa.gov/ssc/data/access/lat/BackgroundModels.html}}. To evaluate the significance of the detection of any source, we used the test statistic $TS = -2\;ln\left( L_{max,0}/L_{max,1}\right)$, where $L_{max,0}$ is log-likelihood value for the null hypothesis and $L_{max,1}$ the log-likelihood for the complete model. The larger the value of the TS, the less likely is $L_{max,0}$. The square root of the TS is approximately equal to the detection significance of a given source. 
 
In the 4FGL DR2 catalogue, the spectrum from $\eta$ Carinae is described with a LogParabola (LP), defined as:

\begin{equation}
		\dfrac{dN}{dE} = N_0\left(\dfrac{E}{E_b}\right) ^{-\left(\alpha+\beta \log{\dfrac{E}{E_b}}\right)}
\end{equation}

where $N_0$ is the normalization flux, $E_b$ is the pivot energy, $\alpha$ the spectral index at $E_b$ and $\beta$ is a curvature parameter.
 
\subsection{Datasets and their analysis}
 
Since $\eta$ Carinae shows two distinct components, we have used different datasets to optimize our analyses. We differentiate between LE and HE \textit{Fermi}-LAT analyses below and above 10 GeV, using the \texttt{FRONT}+\texttt{BACK} event type (evtype=3). We will refer to these hereafter as the LE and HE datasets. But since the large PSF at lower energies could result in source confusion in the Galactic plane and the contamination of our target, we produced a comparison dataset where we only selected the ensemble quartile with the highest quality in the reconstructed direction (evtype=32\footnote{Event types are described in \url{https://fermi.gsfc.nasa.gov/ssc/data/analysis/documentation/Cicerone/Cicerone_Data/LAT_DP.html}}). Hereunder, we will refer to it as the PSF3 dataset, which improves the PSF at $68\%$ from the aforementioned $\sim 5^{\circ}$ to  $\sim 3^{\circ}$ at 100 MeV. The full description of the cuts and details of each dataset is provided in Appendix \ref{A:analysis}. Besides, the model described above has been modified in each particular analysis to assess new background sources (see Section \ref{sec:sources}) and use the appropriate isotropic background according to the instrument response functions (IRFs).

The analysis performed on the LE and PSF3 datasets was similar, fitting the spectrum of the binary with a LP while keeping free the normalisation for all sources within $5^{\circ}$ (i.e. 39 sources) from $\eta$ Carinae present in the 4FGL-DR2, together with all parameters for sources within $1^{\circ}$ of the binary (i.e. 3 sources). On the other hand, we used a PowerLaw (PL) spectrum for the HE dataset, and those distances were reduced to $3^{\circ}$ (i.e. 18 sources) and $0.5^{\circ}$ (i.e. 1 source), respectively. In all datasets, the normalisation of the diffuse components and the spectral index of the Galactic one are also free parameters.

\subsection{Background model extension}\label{sec:sources}

The Carina region is a densely populated, challenging field to characterise with $\gamma$-rays. Multiple sources and strong diffuse emission may distort our fits, and therefore our background model has to be properly updated for our region of interest (ROI). A recent study by \cite{White20} used a CO template of the region to take into account the excesses seen in the residuals. However, we took a different approach. Using the updated 4FGL DR2 already includes a new source within $3^{\circ}$ of $\eta$ Carinae (4FGL J1054.0-5938), but in order to assess the possible excesses present in the residuals, we extended the background model with new sources in an iterative way using the method \texttt{find\_sources} from \texttt{fermipy}. This method has to be used with caution in order to prevent false positives, especially in the Galactic plane. Therefore we employed our different selection datasets to take advantage of their particularities.

\begin{table*}[h]
\caption{Spectral results for $\eta$ Carinae. Energy fluxes are integrated over the energy range selections described in Appendix \ref{A:analysis}.}             
\label{table:etacar}      
\centering                          
\begin{tabular}{c c c c  c}        
\hline\hline      
Parameter & PSF3 (LP) & PSF3 (SBPL) & LE (LP) & HE (PL) \\  
\hline                  
Energy Flux ($10^{-5}$ MeV cm$^{-2}$ s$^{-1}$) & $9.8 \pm 0.2$ & $11.0 \pm 0.5$ & $6.5\pm 0.2$& $1.42 \pm 0.13$ \\
$\Gamma_1$ & -& $2.64 \pm 0.09$ &- & $2.55\pm 0.11$\\
$\Gamma_2$ & -& $1.19 \pm 0.06$ &-  & -\\
$\alpha$ & $2.32 \pm 0.02$& - & $2.24 \pm 0.03$ &  -\\
$\beta$ & $0.17 \pm 0.01$& -& $0.15 \pm 0.03$&  -\\
$\delta$ & -& $1.05 \pm 0.22$ &- &  -\\
$E_b$ (GeV) & $2.1^{a}$ & $0.50 \pm 0.08$& $2.1^{a}$ & -\\

\hline  
\end{tabular}
\tablefoot{$^{a}$Fixed value from 4FGL DR2.}
\label{tab:values}
\end{table*}

The HE is indeed the cleanest dataset in terms of Galactic diffuse contamination, and therefore the most suitable one for searching new (albeit preferentially hard) sources. After our initial fit, we removed all sources detected with less than 2$\sigma$ (TS = 4) in our ROI. This procedure purposefully reduces the degrees of freedom in the fit at the expense of neglecting sub-threshold soft sources. Afterwards, we performed a search for sources on this dataset. We detected two new signals above 4.5$\sigma$, which were added to the model and subsequently re-fitted. In a second iteration, we performed a new search on the PSF3 dataset, using as prior model the new sources in addition to 4FGL DR2. In that case we made our requirements slightly more stringent, demanding at least 5$\sigma$ for a detection. We found 7 source candidates, which were then added to the model and re-fitted. In the last step, all these sources were included in the LE analysis model. Following this procedure, we do not find further significant emission in our residuals coincident with high values in the CO template close to $\eta$ Carinae as in \cite{White20}, since one of the sources is partially coincident with a high CO grammage in the line of sight (see Section \ref{sec:findsources} for a description of the sources and Figure \ref{app:residuals} for the residuals).

\section{Results} \label{sec:results}

\begin{figure}
   \centering
   \includegraphics[width=\columnwidth]{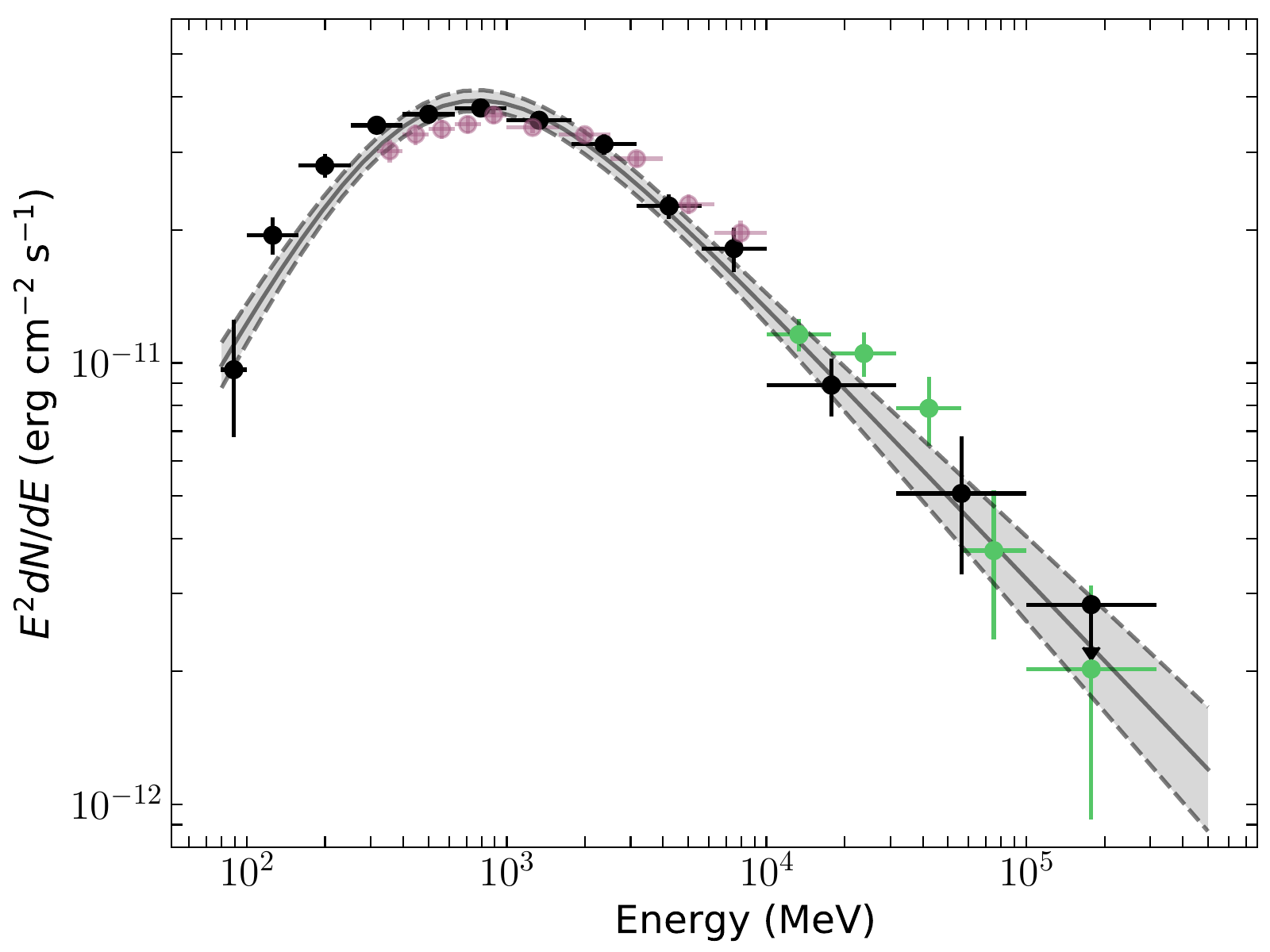}
   
   \caption{Spectral energy distribution for $\eta$ Carinae using the LE (violet), HE (green) and PSF3 (black) datasets. The black line represents the SPBL fit, with the 1$\sigma$ uncertainty shown in grey. }
     
         \label{Fig:etacarsed}
   \end{figure}

Like previous studies \citep{Reitberger15, Balbo17, White20}, we detect $\eta$ Carinae with high significance: 10773 TS, 1017 TS and 7714 TS for the LE, HE and PSF3 datasets, respectively. For the PSF3 dataset, we obtain an energy flux of $\left( 9.8\pm 0.2 \right) \cdot 10^{-5}$  MeV cm$^{-2}$ s$^{-1}$ with $\alpha = 2.32 \pm 0.02$ and $\beta = 0.17\pm 0.01$. For the LE we obtain an energy flux of $\left( 6.5\pm 0.2 \right) \cdot 10^{-5}$ MeV cm$^{-2}$ s$^{-1}$ with $\alpha = 2.24\pm 0.03$ and $\beta = 0.15 \pm 0.03$. Finally, at HE we find $\left( 1.42 \pm 0.13\right) \cdot 10^{-5}$ MeV cm$^{-2}$ s$^{-1}$ with a spectral index $\Gamma = 2.55 \pm 0.10$ (Table \ref{tab:values}). The spectral energy distribution (SED) obtained combining both bands is fully compatible with previous studies, with only minor deviations (Figure \ref{Fig:etacarsed}). The differences might arise from the different cuts employed, the extended dataset or the further updated analysis software.

Similarly to what was done by \cite{Reitberger15}, we also fitted the spectrum obtained in the PSF3 dataset with a smooth broken power law (SBPL):

\begin{equation}
		\dfrac{dN}{dE} = N_0\left(\dfrac{E}{E_0}\right) ^{-\Gamma _1} \left( 1+ \left(\dfrac{E}{E_b}\right) ^{\dfrac{\Gamma _2 - \Gamma _1}{\delta}}  \right)^{-\delta}
\end{equation}

where $E_0$ is the energy normalization (set to 1 GeV), $\delta$ is a curvature parameter and $\Gamma _1$ and $\Gamma _2$ are the spectral indexes above and below the pivot energy $E_b$, respectively. A SBPL model implies $\Gamma_1 = 2.64 \pm 0.09$, $\Gamma_2 = 1.19 \pm 0.06$ and $\delta=1.05\pm0.22$ for the PSF3 dataset (Table \ref{tab:values}). Since for this comparison both hypotheses (LP and SBPL) share a common background model which remains fixed, these are nested \citep[see e.g.][]{Algeri16} and we can compare their likelihood values. We find that the SBPL model is preferred over a LP with $TS = -2 ln \left( L_{LP}/ L_{SBPL}\right)=  93.32$, probably driven from the average emission above 10 GeV.

\subsection{Temporal results}\label{sec:orbit}

\begin{figure}[h]
   \centering
   \includegraphics[width=\columnwidth]{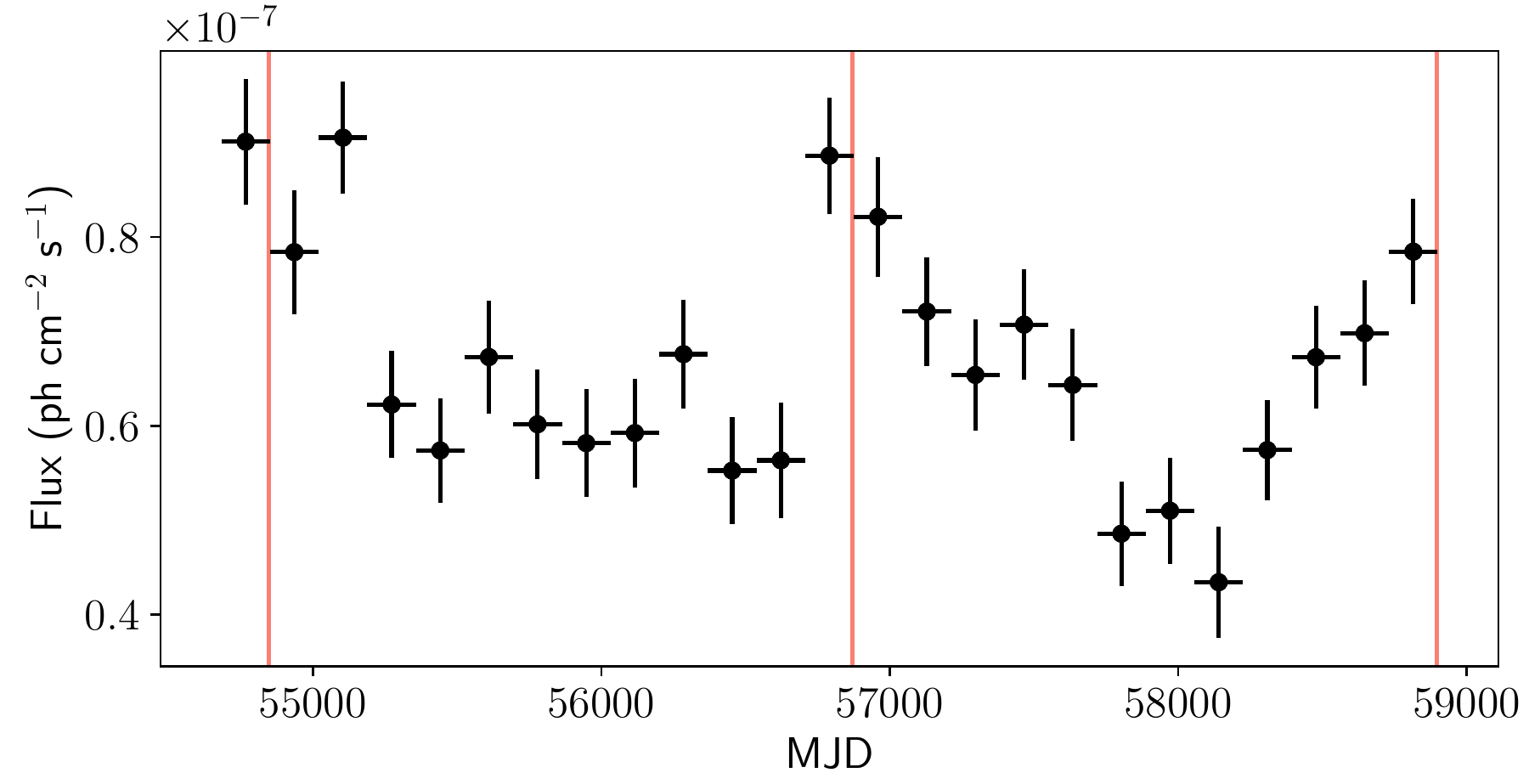}
   
   \caption{Light curve of $\eta$ Carinae using the LE dataset (100~MeV -- 10~GeV), with 12 bins per orbit ($168.67$ days). In red, P2009, P2014 and P2020 periastron passages. At least $5\sigma$ are required per detection in each bin. We do not observe variations of the spectral parameters beyond statistical uncertainties.}
     
         \label{Fig:lcLE}
   \end{figure}

\begin{figure}[h]
   \centering
   \includegraphics[width=\columnwidth]{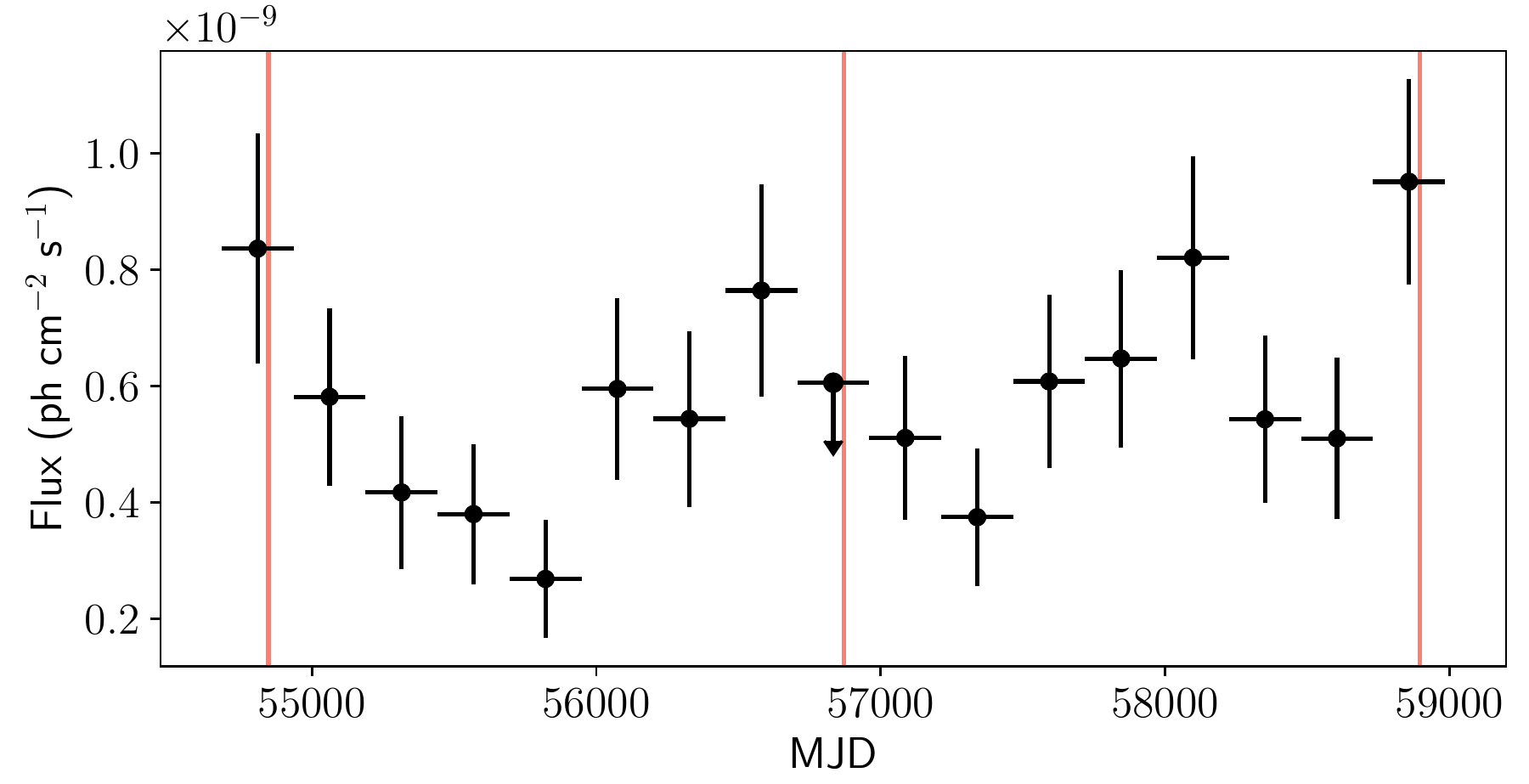}
   
   \caption{Light curve of $\eta$ Carinae using the HE dataset (10~GeV -- 500~GeV), with 8 bins per orbit ($253.0$ days). In red, P2009, P2014 and P2020 periastron passages. At least $5\sigma$ are required per detection in each bin, while a $2\sigma$ upper limit is shown otherwise. We do not observe variations of the spectral index beyond statistical uncertainties.}
     
         \label{Fig:lcHE}
   \end{figure}

In order to study similarities between both orbits, we produced a light curve from the HE and LE datasets described in Section \ref{sec:analysis} using the \texttt{lightcurve} function from \texttt{fermipy}. In each bin we free the parameters from Eta Carinae and the normalization from background sources within $3^{\circ}$, including the diffuse components. We divided each orbit in equally time-spaced bins (Figures \ref{Fig:lcLE} and \ref{Fig:lcHE} for the LE and HE bands, respectively). We observe the $\eta$ Carinae peak around periastron in both bands, and we can distinguish between the first (54848 MJD), the second (56872 MJD) and the third (58896 MJD) periastron as derived from $P = 2024$ days. We will refer to them as P2009, P2014 and P2020, respectively. We employed $P = 2024$ for consistency with earlier works by \cite{Reitberger15} and \cite{Balbo17}; using $P = 2022.7 \pm 1.3$ days as obtained by \cite{Damineli08} does not change the results notably. While at LE the trend is similar for both orbits, the HE component has a different behaviour. The light curve above 10 GeV shows a similar peak for the first and third periastrons, but we observe that the second one occurs several months before periastron. Trying different re-binnings -- by shifting the initial bin or adding more bins -- still produces the peak, and only with a large binning the peak is averaged out (e.g. 6 bins per orbit). Furthermore, a light curve of the least-distant source (Fermi J1042.9-5938) does not show any correlation with the period of $\eta$ Carinae -- particularly not around the peak flux. Note that such an effect would have an impact regardless of the source's spectral index.

\begin{figure*}
   \centering
   \includegraphics[width=0.33\hsize]{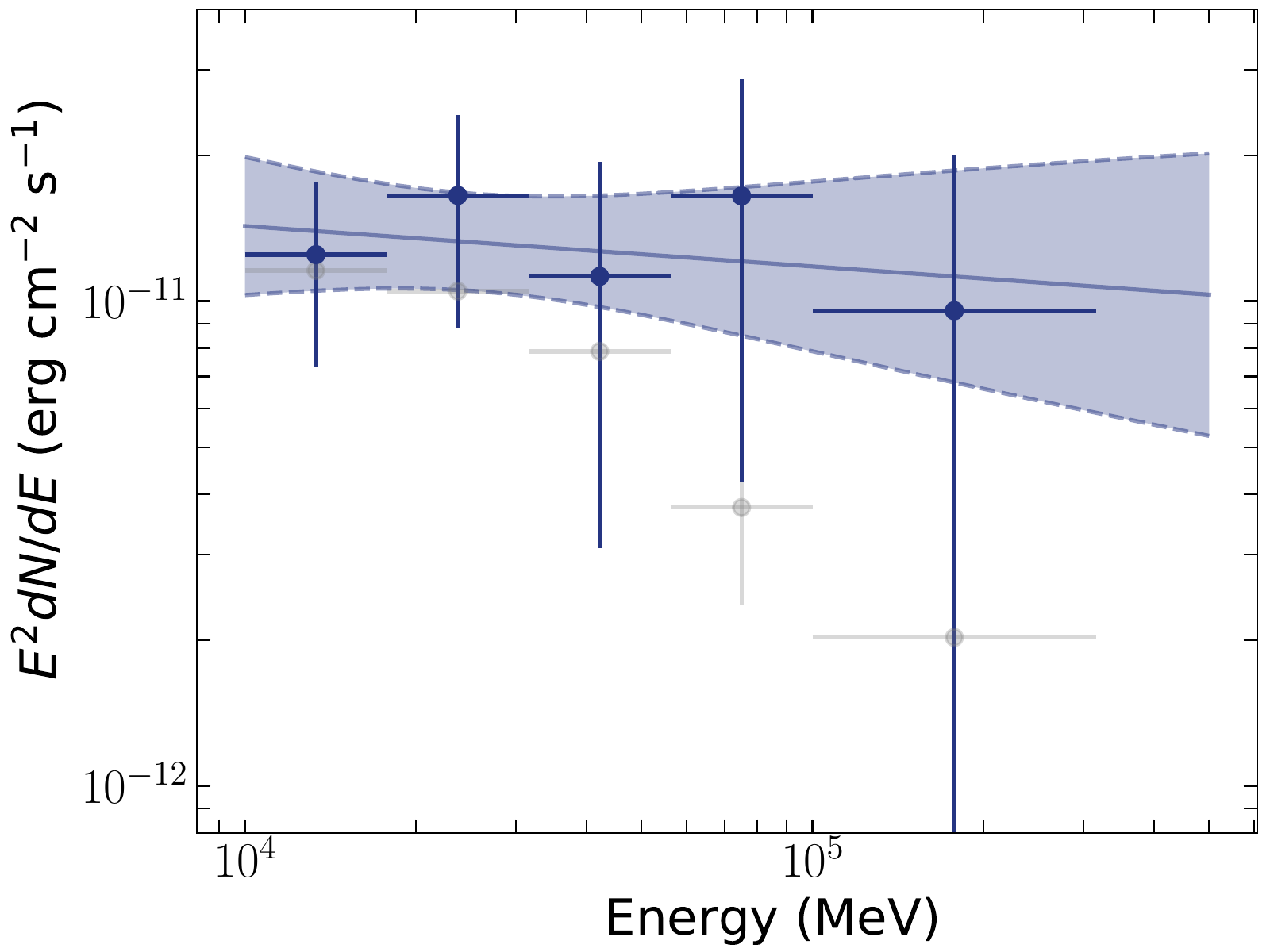}
   \includegraphics[width=0.33\hsize]{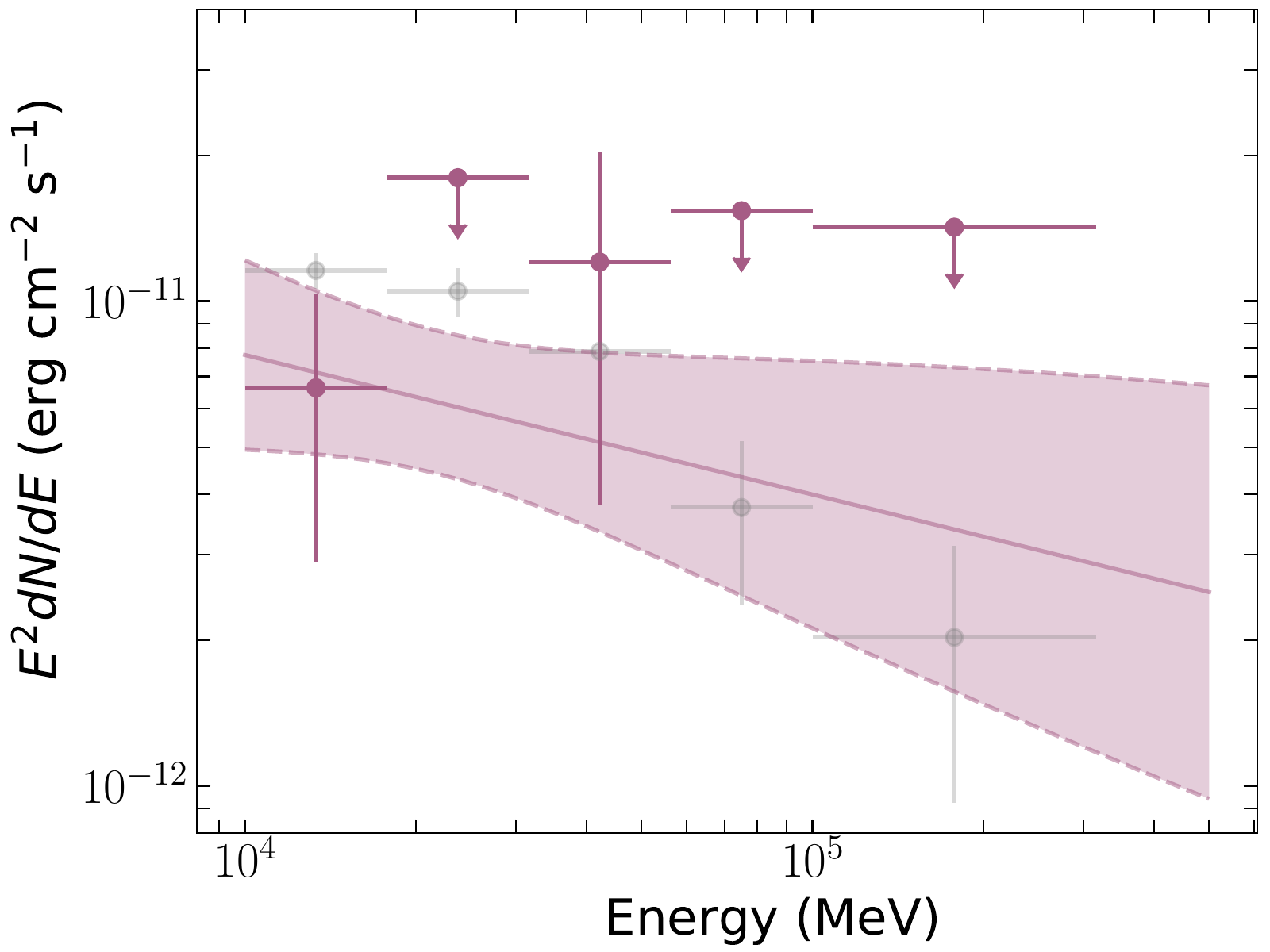}
   \includegraphics[width=0.33\hsize]{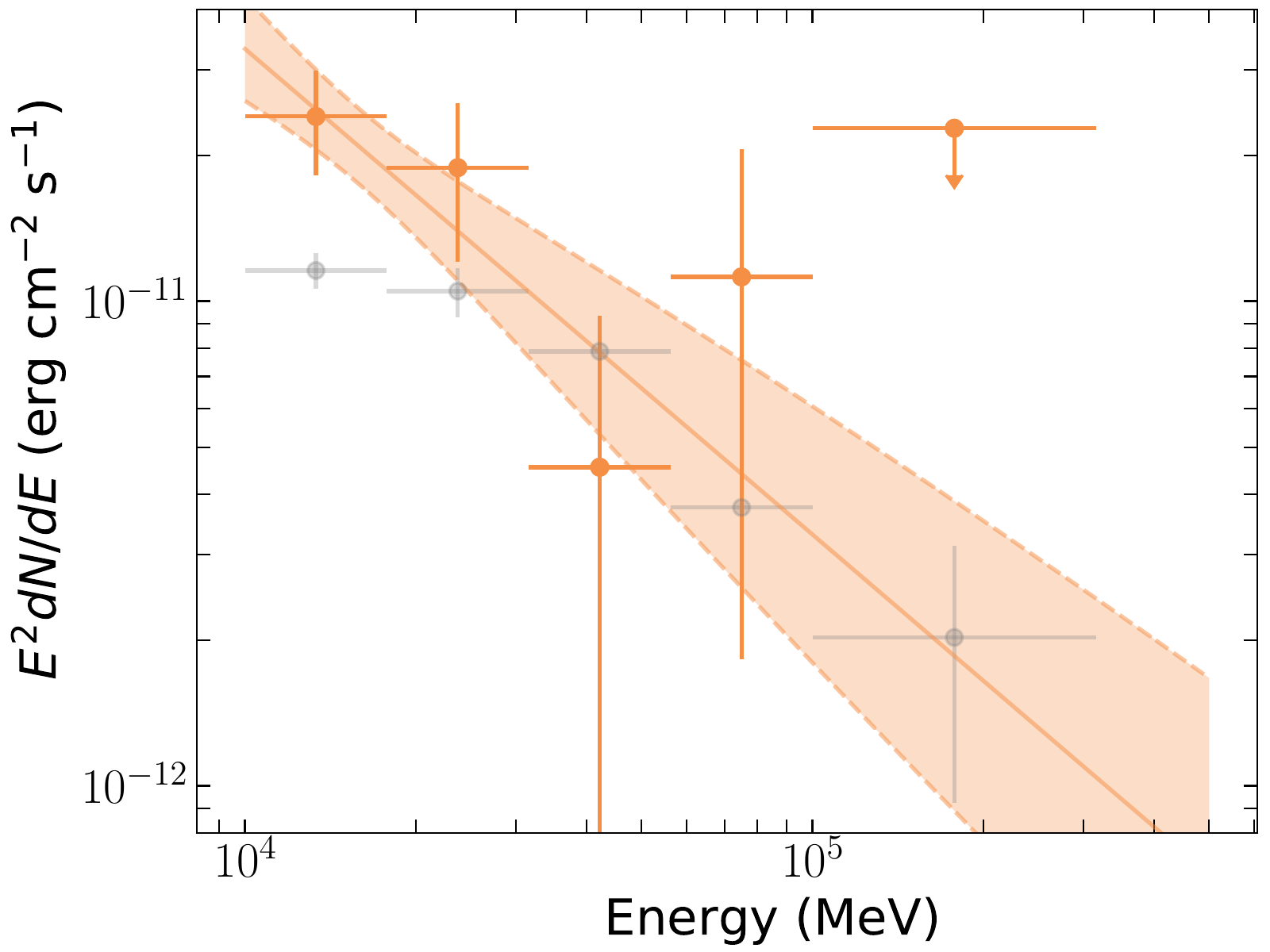}
   \caption{HE SEDs for the three periastrons - P2009 (blue), P2014 (violet) and P2020 (orange) -- observed with Fermi-LAT -- together with the best-fit PL and its $1\sigma$ uncertainty. Each SED is obtained using $\pm 100$ days around each one. For comparison, the overall SED above 10 GeV is shown (grey). }
     
         \label{Fig:Periastrons}
   \end{figure*}

We now study if the spectrum at HE of the peaks observed at periastron varies by producing a SED for each periastron using data from 200 days around them (i.e. $\pm 100$ days). The resulting spectra can be found in Figure \ref{Fig:Periastrons}. Note that this selection is done to ensure sufficient statistics but, given the orbital parameters of $\eta$ Carinae, the distance between both stars does change appreciably during periastron. The results mimic the behaviour observed in the light curve, providing integrated fluxes between 10 GeV and 500 GeV of $(8.12 \pm 2.13) \cdot 10 ^{-10}$ photons cm$^{-2}$ s$^{-1}$ with $TS=71.5$, $(3.69 \pm 1.54) \cdot 10 ^{-10}$ photons cm$^{-2}$ s$^{-1}$ with $TS=21.5$ and $(10.34 \pm 2.08) \cdot 10 ^{-10}$ photons cm$^{-2}$ s$^{-1}$ with $TS=121.9$ for P2009, P2014 and P2020, respectively.

\subsection{Search for nearby sources}\label{sec:findsources}

The $\eta$ Carinae system is in a very densely populated region; thus we studied the possibility of finding new sources nearby the binary and, additionally, searched for $\gamma$-ray emission from nearby CWB systems. 


\subsubsection{Fermi J1042.9-5938 and the nova ASASSN-18fv}\label{sec:newsrcs}

In our analysis of the ROI at HE, we found two new sources: Fermi J1036.1-5934 ($TS = 48.16$; l$ = 286.554 \pm 0.017$, b$ = -1.102 \pm 0.020$) and Fermi J1042.9-5938 ($TS = 35.37$; l$ = 287.345 \pm 0.026$, b$ = -0.713 \pm 0.026$). 

Fermi J1036.1-5934 is spatially coincident with the nova \object{ASASSN-18fv} (or V906 Carinae), which occurred in March 2018 \citep{Jean18}. A light curve of the source confirms that it is only detected during the reported outburst. We modelled it with a PL with an integrated energy flux of $(1.51 \pm 0.16) \cdot 10 ^{-5}$ MeV cm$^{-2}$ s$^{-1}$ and a spectral index of $\Gamma= 2.19 \pm 0.05$. The extended observation time and the more restrictive event cuts are unsuited for the detection of any cut-off as seen by \cite{Aydi20}. On the other hand, no clear counterpart is found for Fermi J1042.9-5938. It partially overlaps with the CO line-of-sight column density as mentioned by \cite{White20}, and it could be related to the source 3FGL J1043.6-5930, not present in the 4FGL DR2. The source is modelled with a PL with an integrated energy flux of $(1.33 \pm 0.21) \cdot 10 ^{-5}$ MeV cm$^{-2}$ s$^{-1}$ and a spectral index of $\Gamma= 2.19 \pm 0.06$. It does not show variability on annual scales, nor any sign of spectral curvature or extension. Both sources were added in the LE analysis model seen in Section \ref{sec:analysis}.

Our search for new sources on the PSF3 dataset also obtained 7 significant detections. However, the detection of these new $\gamma$-ray sources has to be taken with caution at lower energies. Unlike the HE dataset, the PSF3 search is more sensitive to provide excesses caused by the differences between our analysis and the weighted analyses from the original characterisation of the 4FGL \citep{4FGL} or source confusion at low energies. Therefore we consider those excesses only as candidate sources , and we do not explore their nature -- see \citet{Baldini21} for searches of transient sources beyond the 4FGL catalogue.

\subsubsection{Nearby CWBs}\label{sec:cwbs}

The Carina region contains numerous massive stars, some of them in binary systems known to display strong, powerful shocks. A few CWBs from \cite{Becker13} can be found in the vicinity of $\eta$ Carinae: \object{HD~93129A}, \object{HD~93250}, \object{WR~39} and \object{WR~21a}. Besides, two other CWBs with a WR component (\object{WR~22} and \object{WR~25}) are additionally studied given their similarities with $\gamma ^2$ Velorum in terms of stellar components and orbital characteristics \citep{Williams94, Schweickhardt99}. The latter systems are particularly close to Fermi J1042.9-5938, but its localization at HE (where the PSF is smaller than $0.1^{\circ}$) does not favour a tentative association with them. 

\begin{figure}
   \centering
   \includegraphics[width=\columnwidth]{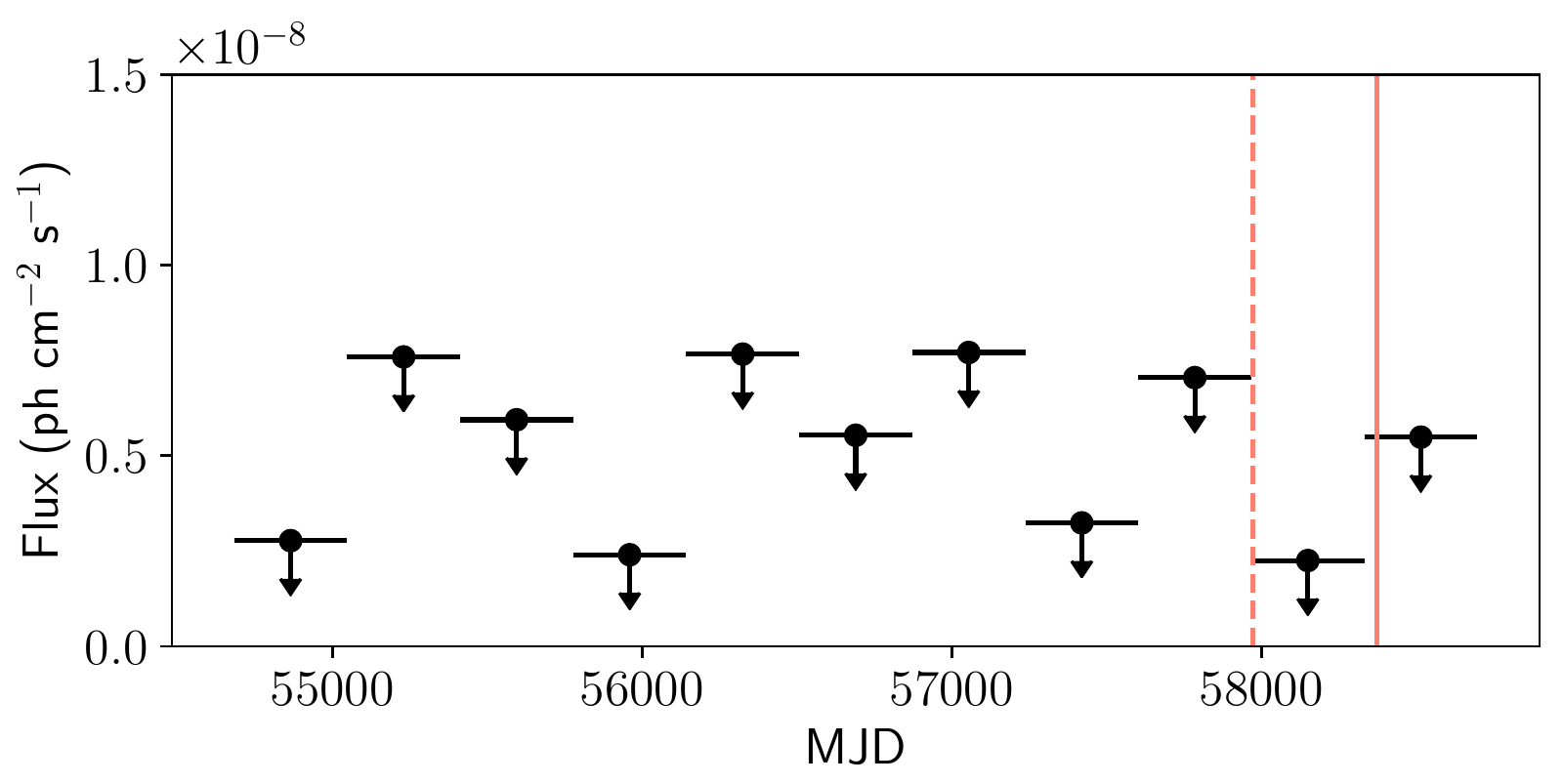}
   \caption{Light curve of HD 93129A with 1 year long bins integrated between 300 MeV and 10 GeV. Upper limits are shown at $2\sigma$ confidence level. The dashed red line (57973 MJD) indicates the periastron passage according to \cite{Maiz-Apellaniz17}, while its updated value from \cite{Palacio20} is marked with a solid red line (58374 MJD). }
         \label{Fig:HD93129A}
   \end{figure}

To evaluate their possible detection, we added a test source using a PL spectrum with $\Gamma = 2$ at the position of each binary, modelling the ROI according to the result from our LE dataset main analysis (see Section \ref{sec:analysis}). We obtained only upper limits, which are summarised in Table \ref{tab:uplims}. Given that the periastron from HD~93129A occurred in late 2017 or early 2018 \citep{Maiz-Apellaniz17}, we produced a light curve for this particular binary. However, no detection above $5\sigma$ is found (see Figure \ref{Fig:HD93129A}).

\begin{table*}
\caption{Upper limits at $95\%$ confidence level obtained other CWBs close to $\eta$ Carinae using the LE dataset. See \cite{Gosset09}, \cite{Becker13}, \citep{Arora19} and references therein.}             
\label{table:cwbs}      
\centering                          
\begin{tabular}{c c c c c c c}        
\hline\hline      
Parameter & HD 93129A & HD 93250 & WR 21a & WR 22 & WR 25 & WR 39 \\  
\hline  
Components & O2f$^\star$ + O3IIIf$^\star$ & O4III + O4III& O3f$^{\star}$/WN6ha + O4& WN7 + O9 & O2.5If$^{\star}$/WN6 + OB & WC7 + ?\\             
Period & $\sim$120 y$^{a}$ & > 100 d & 32.67 d & 80.33 d & 207.85 d & ?\\
Distance & 2900 pc & 2350 pc & 3000 pc & 2700 pc & 2100 pc & 5700 pc\\
\hline
UL (ph cm$^{-2}$ s$^{-1}$) & $1.4\cdot 10^{-9}$& $1.2\cdot 10^{-9}$ & $1.1\cdot 10^{-9}$  & $1.5\cdot 10^{-9}$  & $1.4\cdot 10^{-9}$  & $1.0\cdot 10^{-9}$ \\
TS  & 0.65 & 0.03 & 4.77 & 4.12 & 0.14& 0.69\\

\hline  
\end{tabular}
\tablefoot{$^{a}$Period from \cite{Maiz-Apellaniz17}.}
\label{tab:uplims}
\end{table*}

\section{Discussion} \label{sec:discussion}

These new \textit{Fermi}-LAT results have multiple implications, both from the variability and spectral point of view, and should be compared with previous studies and multi-wavelength data.

\subsection{Orbit-to-orbit variability}

While the variability of the flux at LE does not differ comparing the first and the second orbits, $\eta$ Carinae seems more puzzling above 10 GeV. In the analysis presented by \cite{Balbo17}, the flux at HE during the P2014 periastron did not increase. Our analysis supports that result with a caveat: the flux of $\eta$ Carinae during the P2014 periastron was indeed less bright than the other two -- the flux values around P2014 and P2020 differ at $2.6\sigma$ confidence level -- but the flux was indeed increasing before and its actual peak occurred earlier. Contrarily, the P2020 periastron has a slightly larger flux than P2009. We suggest that such variations might by caused by turbulences and changes in the WCR structure from orbit to orbit, and may indicate that both populations of particles which produce the LE and HE components are accelerated in different regions of the WCR -- e.g. either both sides of the contact discontinuity or at different distances from the apex of the shock.

\begin{figure}
   \centering
   \includegraphics[width=\columnwidth]{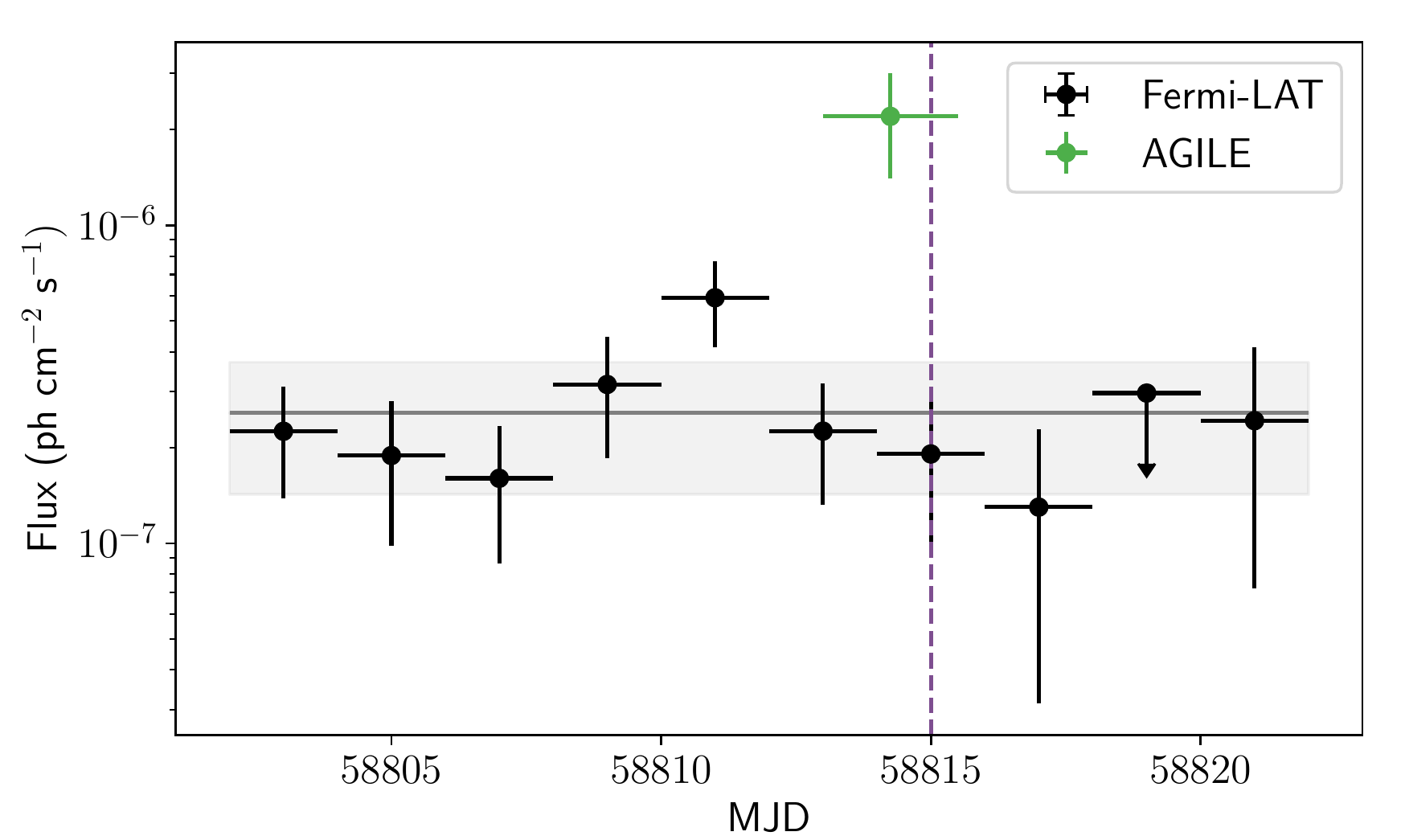}
   \caption{Light curve of $\eta$ Carinae from 15 November to 5 December in 2019. In black, dedicated analysis between 100 MeV and 500 GeV with \textit{Fermi}-LAT (2-day bins, requiring at least $2\sigma$ for detection). The corresponding average flux is represented with a grey line. The $\gamma$-ray flux enhancement reported by \textit{AGILE} is shown in green \citep{Piano20}, which overlaps with the X-ray light curve peak on 28 November 2020, indicated by a violet dashed-line \citep{Corcoran19}.}
     
         \label{Fig:Atel}
   \end{figure}

Especially interesting is the comparison of the present data with the multi-wavelength reports of the P2020 periastron. \textit{NuSTAR} observations of $\eta$ Carinae in the pre-periastron phase reported similar fluxes of non-thermal X-rays compared with previous orbits \citep{Hamaguchi19}. However, the emission was twice as large in the post-periastron phase \citep{Hamaguchi20}. At higher energies, \textit{AGILE} reported a substantially larger $\gamma$-ray flux before the periastron at 4$\sigma$ above 100 MeV \citep{Piano20}, increasing by an order of magnitude with respect to the flux reported in the second \textit{AGILE}-GRID catalogue (2AGL) of $\left(1.81 \pm 0.36 \right) \cdot 10^{-7}$ photons cm$^{-2}$ s$^{-1}$ \citep{Bulgarelli19}. Contrarily, a \textit{Fermi}-LAT analysis above 100 MeV during the same dates provides a flux of $\left( 1.42 \pm 0.78 \right) \cdot 10^{-7}$ photons cm$^{-2}$ s$^{-1}$ ($TS=2.4$), thus consistent with the 2AGL result. Besides, a light curve around those dates does not seem to support such rise of the $\gamma$-ray flux (see Figure \ref{Fig:Atel}). This analysis only differs with the LE analysis described in Section \ref{sec:analysis} in the energy range, dates and background model (only includes 4FGL DR2). We also explored the possible impact of the different observing modes of \textit{Fermi} during those dates\footnote{See \url{https://fermi.gsfc.nasa.gov/ssc/observations/timeline/posting/ao12/}} on the exposure. Indeed, the observing mode of \textit{Fermi} varies for that time interval. However, we find that the exposure varies less than a factor 2 between the different bins shown in Figure \ref{Fig:Atel} -- thus being an unlikely origin for the discrepancy, even if it was not properly assessed in our likelihood analysis.

\subsection{On the origin of the LE component}\label{sec:LE-model}

Although the HE component is commonly believed to have a hadronic origin, the LE one remains the major focus of discussion around $\eta$ Carinae. Both leptonic \citep{Farnier11, Balbo17} and hadronic \citep{Ohm15} models have been proposed to explain its origin with inconclusive results. To distinguish between both scenarios, \cite{White20} used the low energy regime of \textit{Fermi}-LAT to search for a signature of the $\pi ^0$-bump. According to their results, the spectrum could not be reproduced with a leptonic model. However, the same authors performed multiple analyses and found that a single PL connecting both hard X-rays and $\gamma$-rays could not be discarded at $68\%$ confidence level. The spectral shape at lower energies (especially below 100 MeV) is significantly affected by source confusion due to the large PSF and strong Earth limb contamination, impacting the results in typical analyses with \textit{Fermi}-LAT and therefore requiring alternative analyses \citep{Giacomo18}. Our analysis with more stringent cuts on the PSF reconstruction provides a more conservative reference and does not confirm a significant sudden drop in the spectrum at 100 MeV. Using an SBPL model, the preferred spectral index is harder than for a single PL X-ray/$\gamma$-ray connection with $\Gamma \approx 1.65$ \citep{Hamaguchi18} suggesting that a single PL scenario is disfavoured. But, incidentally, the indication of lower flux observed at 80--100 MeV may also have its origin in an over-subtraction of events due to a large normalisation of the diffuse components. Furthermore, note that the 3FGL all-sky analysis demonstrated how, on average, the systematic uncertainty on the flux due to the Galactic diffuse emission resembles the statistical error \citep{3FGL}.

\subsection{The population of CWBs at high energies}
\label{sec:pop}

The non-detections of CWBs in HE $\gamma$-rays \citep{Werner13, Pshirkov16} lead to strict constraints of their capabilities as particle accelerators. The number of such systems detected at $\gamma$-rays is still scarce, with the only confirmed cases $\eta$ Carinae itself and $\gamma ^2$ Velorum \citep{Pshirkov16, Marti-Devesa20}\footnote{We also note the tentative association of a new CWB in the 4FGL DR2 \citep{Ballet20}. The $\gamma$-ray source 4FGL J1820.4-1609c is found to be compatible with \object{CEN~1} (also known as Kleinmann's star), a trapezium system formed by two O+O binaries (CEN~1a and CEN~1b) showing variable non-thermal radio emission from one of its components \citep{Rodriguez12}. However, studying this system in detail is beyond the scope of this paper.}.

Both confirmed cases behave differently, showing that the $\gamma$-ray emissivity is highly dependent on the particularities of each system. The case of $\gamma ^2$ Velorum showed HE emission during apastron, with no detection of radio synchrotron \citep{Benaglia19, Marti-Devesa20}. This is unexpected because classical models estimate $L_{\gamma} \sim 1/d$, where $d$ is the distance between both components of a binary. This result invites us to reconsider emission coming from other systems, like WR 22, with a period of 80 days and all the particularities of other CWBs, but not detected at non-thermal radio frequencies \citep{Parkin11, Becker13}. Although it is similar to $\gamma ^2$ Velorum, WR 22 is located at a distance of $\sim 2.7$ kpc \citep{Gosset09}. Therefore a comparable $\gamma$-ray luminosity to $\gamma ^2$ Velorum would provide fluxes below the sensitivity limit of \textit{Fermi}-LAT. Another example of a CWB without clear non-thermal radio emission in the Carina region with similar characteristics is WR 25 \citep{Arora19}, which also remains undetected in our study. Since both binaries were not included in previous searches, we constrain the $\gamma$-ray emission of these CWBs with first upper limits.
  
On the other side, HD~93129A is a more classical CWB which has been studied in detail recently. Its WCR was resolved by \cite{Benaglia15} at radio frequencies, showing that it was indeed a CWB system with synchrotron emission. This binary is composed by the earliest-type O stars in the CWB catalogue by \cite{Becker13}, in a very wide orbit with eccentricity $e>0.95$ and period of $\sim120$ years \citep{Maiz-Apellaniz17}. Its periastron passage was estimated to occur around the early months of 2018, giving a unique opportunity to study very close periastron conditions in CWBs. Non-thermal emission was predicted during its periastron passage at high energies  \citep{Palacio16}, but no X-ray non-thermal component nor $\gamma$-rays have been detected using \textit{NuSTAR} nor \textit{AGILE} \citep{Palacio20}. Our upper limits further constrain its putative emission between 300 MeV and 10 GeV, with an upper limit during the periastron passage of $ 5.49 \cdot 10^{-9}$ photons cm$^{-2}$ s$^{-1}$ (Figure \ref{Fig:HD93129A}). Although this limit is almost two orders of magnitude lower than the previous constraints from \textit{AGILE}, it does not conflict with the updated emission model by \cite{Palacio20} considering its non-detection at hard X-rays.

\section{Conclusion} 
\label{sec:summary}

In this work, we presented a comprehensive study on the second full orbit of $\eta$ Carinae and its third periastron as seen by the \textit{Fermi}-LAT. Our results also hint that the recent periastron passage in February 2020 was the brightest observed with $\gamma$-rays. We found evidence of orbit to orbit variability in this system above 10 GeV, suggesting that the transport of particles in the WCR might be affected by different turbulences in each orbit perturbing such structure. To study the origin of the LE component, we used a stringent cut on the reconstruction quality of the PSF, but we can not confirm a significant $\pi^0$-bump. Complementarily, we searched for new sources in the Carina region not included in the 4FGL catalogue and found two sources above 10 GeV (one associated with the nova V906 Carinae) and seven candidate sources at lower energies. Unfortunately, no emission has been found coincident with other massive CWBs present in the Carina region. 

In short, the complicated behaviour observed prevents any simplistic considerations of particle acceleration in $\eta$ Carinae and other CWBs. A future work will explore the modelling of $\eta$ Carinae with magnetohydrodynamical simulations to explain the observed phenomena reported in this study.

\begin{acknowledgements}
The Fermi LAT Collaboration acknowledges generous ongoing support from a number of agencies and institutes that have supported both the development and the operation of the LAT as well as scientific data analysis. These include the National Aeronautics and Space Administration and the Department of Energy in the United States, the Commissariat \`{a} l'Energie Atomique and the Centre National de la Recherche Scientifique / Institut National de Physique Nucl\'{e}aire et de Physique des Particules in France, the Agenzia Spaziale Italiana and the Istituto Nazionale di Fisica Nucleare in Italy, the Ministry of Education, Culture, Sports, Science and Technology (MEXT), High Energy Accelerator Research Organization (KEK) and Japan Aerospace Exploration Agency (JAXA) in Japan, and the K. A. Wallenberg Foundation, the Swedish Research Council and the Swedish National Space Board in Sweden. Additional support for science analysis during the operations phase from the following agencies is also gratefully acknowledged: the Istituto Nazionale di Astrofisica in Italy and the Centre National d'Etudes Spatiales in France. This work performed in part under DOE Contract DE-AC02-76SF00515. \\

This work uses DSS2 images accessed via \url{http://archive.eso.org/dss/dss}. Southern hemisphere DSS2 data is based on photographic data obtained using The UK Schmidt Telescope. The UK Schmidt Telescope was operated by the Royal Observatory Edinburgh, with funding from the UK Science and Engineering Research Council, until 1988 June, and thereafter by the Anglo-Australian Observatory. Original plate material is copyright (c) of the Royal Observatory Edinburgh and the Anglo-Australian Observatory. The plates were processed into the present compressed digital form with their permission. The Digitized Sky Survey was produced at the Space Telescope Science Institute under US Government grant NAG W-2166.
\end{acknowledgements}

%
%

\bibliographystyle{aa}
\bibliography{sample.bib}

\begin{appendix}
\section{\textit{Fermi}-LAT analysis} \label{A:analysis}

\begin{table}[h]
\caption{Description of the LE, HE and PSF3 datasets}             
\label{table:1}      
\centering                          
\begin{tabular}{c c c c}        
\hline\hline                 
Parameter & LE & HE & PSF3 \\    
\hline                        
  Data & P8R3 & P8R3 & P8R3 \\  
  IRFS & P8R3\_SOURCE\_V2 & P8R3\_SOURCE\_V2 & P8R3\_SOURCE\_V2 \\     
  ROI & $20^{\circ} \times 20^{\circ} $ & $8^{\circ} \times 8^{\circ} $ & $20^{\circ} \times 20^{\circ} $ \\
  binning & $0.1^{\circ}$ & $0.1^{\circ}$ & $0.1^{\circ}$\\
  N$_{ebins}$ & $10$ & $8$ & $10$\\
  E$_{min}$ & $300$ MeV & $10$ GeV & $80$ MeV \\
  E$_{max}$ & $10$ GeV & $500$ GeV & $500$ GeV\\
  z$_{max}$ & $90^{\circ}$ & $105^{\circ}$  & $90^{\circ}$  \\ 
  evclass & $128$ & $128$  & $128$ \\ 
  evtype & $3$  & $3$  & $32$ \\ 
  edisp & True$^{a}$ & True$^{a}$ & True$^{a}$\\
  edisp binning & $-2$ & $-2$ & $-2$ \\
  DATA\_QUAL & 1 & 1 & 1 \\ 
  LAT\_CONFIG & 1 & 1 & 1 \\ 
  ROI model width & $30^{\circ} \times 30^{\circ} $  & $12^{\circ} \times 12^{\circ} $ & $30^{\circ} \times 30^{\circ} $\\
\hline                                   
\end{tabular}
\tablefoot{$^{a}$Except for the isotropic background}
\end{table}

\begin{figure}[H]
   \centering
   \includegraphics[width=\columnwidth]{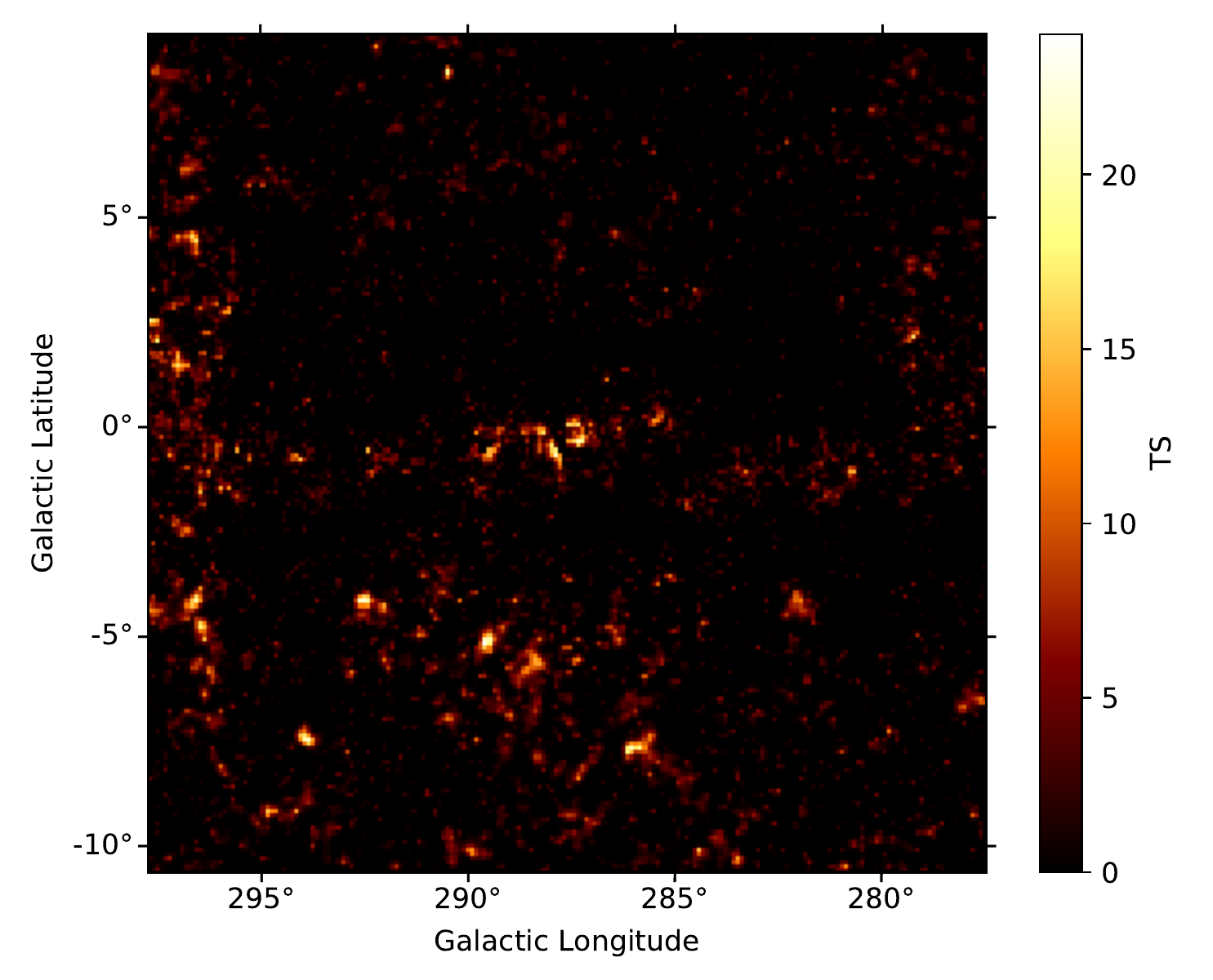}
   
   \caption{TS map of the ROI using the PSF3 dataset, after removing all residual sources as described in Section \ref{sec:sources}.}
     
         \label{app:residuals}
\end{figure}

\end{appendix}
\end{document}